# Experimental percolation studies of two-dimensional honeycomb lattice: $Li_2Mn_{1-x}Ti_xO_3$

Sanghyun Lee[1,2,3], Junghwan Park[2,3], Jiyeon Kim[2], Kun-Pyo Hong[1,4], Youngmi Song[5], and Je-Geun Park[1,2,6#]

[1]*Center for Correlated Electron Systems, Institute for Basic Science (IBS), Seoul 151-747, Korea*

[2]*Center for Strongly Correlated Materials Research, Seoul National University, Seoul 151-742, Korea*

[3]*Department of Physics, SungKyunKwan University, Suwon 440-746, Korea*

[4]*Neutron Science Division, Korea Atomic Energy Research Institute, Daejeon 305-353, Korea*

[5]*National Center for Inter-University Research Facilities, Seoul National University, Seoul 151-742, Korea*

[6]*Department of Physics & Astronomy, Seoul National University, Seoul 151-742, Korea*

## Abstract

$Li_2MnO_3$ with a *S*=3/2 two-dimensional Mn honeycomb lattice has a Neel-type antiferromagnetic transition at $T_N$=36 K with a broad maximum in the magnetic susceptibility at $T_M$=48 K. We have investigated site percolation effects by replacing Mn with nonmagnetic Ti, and completed a full phase diagram of $Li_2Mn_{1-x}Ti_xO_3$ solid solution systems to find that the antiferromagnetic transition is suppressed continuously without a clear sign of changes in the Neel-type antiferromagnetic structure. The magnetic ordering eventually disappears at a critical concentration of $x_c$=0.7. This experimental observation is consistent with percolation theories for a honeycomb lattice when one considers up to 3$^{rd}$ nearest-neighbor interactions. This study highlights the importance of interaction beyond nearest neighbors even for Mn element with relative localized 3d electrons in the honeycomb lattice.

# 1. Introduction

Over the past few years, there has been growing interest in the honeycomb lattice, where each site has only three nearest neighbors, the smallest number possible for regular two dimensional systems. For example, honeycomb lattices consisting of magnetic ions have been considered in the context of the Kitaev model [1], where frustrated, directional anisotropic nearest neighbor interactions yield a spin liquid ground state. Quasiparticles produced in such systems are proposed as a candidate for quantum computing. This model was subsequently further developed for cases with a strong spin-orbit coupling and extended to include a Heisenberg term, i.e. a Kitaev-Heisenberg model, with rich ground states theoretically predicted [2].

$A_2TMO_3$ (A=Li and Na, TM=3d, 4d and 5d transition metal elements) forms in one of the three crystal structures ($P2_1/m$, $C2/m$, and $C2/c$) with TM forming a honeycomb lattice, which is then sandwiched by $LiO_6$ layers [3]. Here, we have a flexibility in the material design of controlling spin value $S$, bandwidth $W$ and coulomb interaction $U$ by varying TM elements from 3d to 5d electrons with different degrees of localization and spin-orbit coupling (SOC). Because of the edge-sharing $TMO_6$ octahedra arrangement, d-d direct exchange and TM-O-TM 90° super-exchange produce antiferromagnetic and ferromagnetic interactions, respectively, which then compete with one another [4].

$Li_2MnO_3$ has a Mn honeycomb lattice with S=3/2 [3]. X-ray (XRD) and neutron diffraction studies show very small site disorders between Mn and Li elements, so confirming a good crystalline nature of the honeycomb lattice. Resistivity of $Li_2MnO_3$ follows the usual Arrhenius law $\rho(T) \propto \exp(\frac{\Delta}{T})$, while $Li_2RhO_3$ and $Na_2IrO_3$ are reported to exhibit three-dimensional variable-range-hoping behavior [5, 6]. We estimated a gap energy $\Delta$ of $Li_2MnO_3$ to be about 600~700 meV, as compared with $\Delta$=78 meV for $Li_2RhO_3$ and $\Delta$=340~400 meV for $Na_2IrO_3$ [3, 5, 7, 8]. Resistivity measured on a single crystal $Li_2MnO_3$ sample is strongly anisotropic so that the out-of-plane resistivity is 10 times larger than the in-plane resistivity [3].

Since Mn has a smaller SOC value than its counterparts of 4d and 5d elements, $Li_2MnO_3$ is close to a classical Néel-type antiferromagnet in the Kitaev-Heisenberg model [2, 9]. Thus it may not be surprising that $Li_2MnO_3$ undergoes an antiferromagnetic ordering at $T_N$=36 K after a broad maximum in the magnetic susceptibility with a peak at $T_M$=48 K. At the same time, heat capacity measurement shows a long tail above the transition temperature, and as much as 35% of the total magnetic entropy is released above $T_N$. Some of the unusual heat capacity behavior is believed to arise from the two-dimensional nature of the honeycomb lattice. Our previous neutron diffraction studies of $Li_2MnO_3$ also showed that the Mn spins order in a Néel-type antiferromagnetic structure with Mn spins aligned perpendicular to the ab plane [3]. It is interesting to note that $Bi_3Mn_4O_{12}(NO_3)$ with Mn bilayer honeycomb lattice doesn't have antiferromagnetic long-range order down to 0.4 K [10, 11]. The nearest Mn-Mn distance on honeycomb lattice is ~2.84 Å for $Li_2MnO_3$ whereas it is ~2.87 Å for $Bi_3Mn_4O_{12}(NO_3)$ [3, 10]. The shorter distance of nearest Mn and Mn distance in $Li_2MnO_3$ as compared to $Bi_3Mn_4O_{12}(NO_3)$ can give rise to more overlap of 3d orbitals so strengthens the antiferromagnetic interaction. On the other hand, a large next-nearest neighbor interaction ($J_2$) compared with the nearest neighbor interaction ($J_1$) is crucial for the disordered magnetic ground state for $Bi_3Mn_4O_{12}(NO_3)$ as discussed in Ref. 11: the critical value is estimated to be $J_2/J_1 \sim 1/6$. Thus it is an extremely interesting question how the two materials with the small difference in the Mn-Mn distance can have such a large different ratio of $J_2/J_1$ and so disparate ground states.

In order to better understand the physical properties of $A_2TMO_3$ in general and those of $Li_2MnO_3$ more specifically, it is important for us to know how the magnetic phase transition evolves upon doping. Traditionally, doping experiment and so studies of percolation effects have played an important role in improving our understanding of the generic behavior of magnetic materials [12]. Furthermore, we note that there have been relatively few experimental studies of the percolation effects on systems [13-15] with a honeycomb lattice, less still a complete phase diagram of doping experiments covering an entire phase diagram from one magnetic end compound to another nonmagnetic end compound.

In this paper, we investigated the structural and magnetic properties of $Li_2Mn_{1-x}Ti_xO_3$ solid solution between $Li_2MnO_3$ (Mn $S$=3/2, C2/m) [3] and $Li_2TiO_3$ (Ti $S$=0, C2/c) [16]. Because of the similar crystal structures of both end compounds, although not exactly the same, we can follow the doping dependence of the magnetic transition over almost the entire composition range except for a small window of a miscibility gap. The critical concentration of $x_c$=0.7, where the magnetic transition disappears altogether, shows good agreement with theoretical studies, and our studies provide a rare opportunity of full experimental percolation studies on a honeycomb lattice.

## 2. Experimental details

We prepared several $Li_2Mn_{1-x}Ti_xO_3$ samples: powder samples with x=0, 0.1, 0.25, 0.4, 0.5, 0.64, 0.68, 0.72, 0.75, 0.8, 0.9, and 1.0 by using a solid-state reaction method and single crystals with x=0.0, 0.26(4), and 0.46(2) by employing a flux method as reported in Ref. 3.

We checked the quality and crystal structure of all our samples by using XRD (MiniFlex Ⅱ, Rigaku) and EPMA (JXA-8900R, JEOL). We used two commercial set-ups to measure the magnetization from 1.8 to 300 K: for single crystals a SQUID magnetometer (MPMS-5XL, Quantum Design) with magnetic field 30 Oe applied along the out-of-plane and in-plane directions of the honeycomb lattice; for powder samples a VSM magnetometer (PPMS-14, Quantum Design) with applied magnetic field of 30 Oe. We also conducted powder neutron diffraction studies of $Li_2Mn_{1-x}Ti_xO_3$ for x=0.25 and 0.5 between $2\theta$=10~160°, using a wavelength of $\lambda$=1.8343 Å of Ge (331) monochromater at HRPD beamline of HANARO, Korea. We employed Rietveld program *Fullprof* to analyze the crystal and magnetic structure of both XRD and neutron diffraction data [17].

## 3. Results and analysis

$Li_2MnO_3$ (C2/m) and $Li_2TiO_3$ (C2/c) form in the honeycomb lattice with their respective crystal structure as shown in Fig. 1. Each Mn and Ti elements sit at the center of the oxygen octahedron that shares the edges with the neighboring octahedra. In both structures, the honeycomb lattices of Mn/Ti are sandwiched by the $LiO_6$ layers. For the C2/m crystal structure, the two-dimensional honeycomb lattices are stacked along the c-axis after being shifted horizontally along the a-axis. On the other hand, the two-dimensional honeycomb lattices are stacked along the c-axis for the C2/c crystal structure, but shifted horizontally along both a and b-axes. Our analysis of room-temperature XRD data produces the following set of lattice constants: for $Li_2MnO_3$ a=4.9246(3), b=8.5192(5), c=5.0220(3) Å, β=109.397(3)°; for $Li_2TiO_3$ a=5.0585(2), b=8.7780(4), c=9.7452(4) Å, β=100.183(3)°. Note that the unit cell of $Li_2TiO_3$ with the C2/c structure is twice larger than that of $Li_2MnO_3$ with the C2/m structure.

Because of the similarity of the two end structures, we can dope the TM site rather easily with a very narrow miscibility gap. One passing note on the sintering temperature: $Li_2MnO_3$ synthesized below 800 ℃ lower than our optimum conditions showed a broad background at 2θ=20~30°, which is indicative of Li and Mn disorder in the honeycomb lattice as well as stacking faults along the c-axis [18, 19]. As one can see in Fig. 2, all our samples show a rather flat background in the 2θ range of interest attesting the good crystallinity of all our samples. With increasing Ti composition, there are continuous shifts in the peak position of all Bragg peaks with the C2/m crystal structure until it reaches x=0.64. Then there is a narrow miscibility gap with the C2/m and C2/c structures coexisting in the regions of x=0.68~0.75. For x > 0.75, the system adopts the C2/c structure of $Li_2TiO_3$. For example, we marked the C2/c superlattice peaks in the data for x=0.8 by asterisks in the figures. As the C2/c structure has a twice larger formula per unit than the C2/m structure, we have divided the unit cell of the C2/c structure by a factor 2 to compare it with that of the C2/m structure in Fig. 6. It is clear in the figure that the unit cell volume of $Li_2Mn_{1-x}Ti_xO_3$ follows the Vegard's law (dashed line) rather well. Our XRD data confirm that $Li_2Mn_{1-x}Ti_xO_3$ is an ideal system to study a percolation problem for a honeycomb lattice with a wide solid solution region.

Using both polycrystals and single crystals, we studied how the magnetic susceptibility evolves upon doping nonmagnetic Ti at the Mn site as shown in Fig. 3. We measured both out-of-plane ($\chi_{OP}$) and in-plane magnetic susceptibility ($\chi_{IP}$) of single crystals with magnetic field 30 Oe applied perpendicular and parallel to the honeycomb plane as shown in the main body of Fig. 3. We also plotted data for four representative powder samples in the inset for better clarity although we measured all the power samples: the summary of the data is given in the bottom figure of Fig. 3 and Fig. 6. The single crystal data show that the magnetic easy axis is still along the c*-axis or perpendicular to the honeycomb lattice for all three samples. It is also noticeable that the broad maximum becomes almost absent from the magnetic susceptibility data for x=0.26. We note that similar suppression of a broad maximum in the susceptibility was also reported in $Mn_pZn_{1-p}PS_3$ [13] and $K_2Mn_pMg_{1-p}F_4$ [20].

Using the temperature dependence of the magnetization data measured up to 300 K, we estimated Curie-Weiss temperature $\theta_{CW}$ for all our samples over a temperature range from 100 to 300 K: the inset in Fig. 3 shows the 1/χ vs T plot for x=0.64, where the Curie-Weiss fitting is valid over a much larger temperature range. When we combine all our data collected on single crystals as well as powder samples, a clear pattern emerges that the Curie-Weiss temperature $\theta_{CW}$ is continuously reduced with Ti doping and even change its sign at around x=0.8, indicating that the dominant magnetic interaction changes from antiferromagnetic for x < 0.8 to ferromagnetic for x > 0.8 (see the bottom figure of Fig. 3). Interestingly, this sign change occurs at the critical composition of $x_c$=0.7 where the crystal structure change from C2/m to C2/c. The gradual decrease of the Curie-Weiss temperature with Ti doping is most likely to be due to the fact that there is a smaller number of neighboring Mn moments for doped materials as compared to $Li_2MnO_3$. However, it is rather unexpected to see the change in the sign of $\theta_{CW}$ when it changes its crystal structure from one honeycomb structure to another. The sign change in $\theta_{CW}$ can be natrually understood in a way that the exchange paths mediating the magnetic interaction in the C2/c structure is different from that in the C2/m structure because of the different stacking along the c-axis as we discussed in the introduction. It is an interesting manifestation of the two different stacking on the magnetic properties of the honeycomb lattice: similar effects of different stacking attracted considerable interest in graphene [21].

In order to study whether the magnetic structure of $Li_2MnO_3$ changes with Ti doping, we carried out neutron diffraction studies and present representative data in Fig. 4 for three samples x=0, 0.25 and 0.5 with the summary given in Table 1. It is clear even in the raw data that the position of the magnetic Bragg peaks appears almost unchanged with doping apart from weaker intensity seen in the data for the doped samples. Indeed, our Rietveld analysis shows that the magnetic peaks of the two doped samples can be well explained by a $\Gamma_{2u}$ magnetic structure model, which we found previously for $Li_2MnO_3$ with a magnetic propagating vector $k_m$=(0 0 0.5) [3]. In this magnetic structure, Mn moments are aligned antiferromagnetically within the same honeycomb lattices and they are also antiferromagnetically coupled along the c-axis with a magnetic propagating vector of $k_m$=(0 0 0.5). The temperature dependence of the perpendicular (OP) and parallel (IP) spin components to the ab-plane are shown in Fig. 5 with the lines representing

$$M(T) = M_0 \left( \frac{T_N - T}{T_N} \right)^\beta$$

using β=0.28. We caution here that the critical exponent β value we used is only intended to serve as a fitting parameter: a more detailed examination of the critical exponent was carried out in our previous studies of single crystal $Li_2MnO_3$ [3]. We note that our powder neutron diffraction data are also consistent with the magnetic easy-axis obtained from the magnetic susceptibility measurement of single crystals as shown in Fig. 3. We would like to comment that the $C_x$ magnetic model with moments aligned along the a-axis proposed in Ref. 22 is at variance with all our data. It is also interesting to note that in another honeycomb lattice of $Mn_pZn_{1-p}PS_3$ the magnetic structure was reported to change its spin direction by nonmagnetic Zn doping [13]. We summarize the magnetic data in the full phase diagram of Fig. 6.

## 4. Discussion and Summary

The honeycomb lattice has two sub-lattices, so it is not geometrically frustrated if only nearest-neighbor interaction $J_1$ is considered. However, when more extended interactions are taken into account up to 3$^{rd}$ nearest-neighbor interaction $J_3$, recent theories suggest that various magnetic ground states should be realized for a honeycomb lattice: *i.e.* classical Néel type ordering, stripy, zigzag, and spiral magnetic ordering [9].

For $Li_2MnO_3$, every $MnO_6$ octahedron shares their edge with neighboring octahedra and two oxygen atoms are located off the Mn-Mn plane. This geometrical arrangement allows Mn and Mn orbitals to directly overlap with each other unlike in the usual case of localized Mn 3d orbtials in perovskites, and thus gives rise to antiferromagnetic d-d direct exchange interaction while Mn-O-Mn 90° super-exchange gives ferromagnetic interactions. That the magnetic structure of the $\Gamma_{2u}$ representation has spins aligned antiferromagnetically on the honeycomb lattice suggests that the d-d direct exchange interaction is relatively stronger than the Mn-O-Mn 90° super-exchange interaction. The fact that the $\Gamma_{2u}$ magnetic structure survives all the way to the critical composition implies that this hierarchy of energy scales appears to be unchanged with Ti doping. This can be then taken as a sign that even for materials like $Li_2MnO_3$ with supposedly localized 3d orbital exchange interactions beyond nearest-neighbor interaction are important. On the other hand, Ti doping reduces the overlap of Mn d orbitals above the critical composition, causing the Curie-Weiss temperature $\theta_{CW}$ to change its sign from negative to positive near x=0.8. We think that this lesson on the importance of direct exchange interactions in $Li_2(Mn,Ti)O_3$ may be of interest to other transition metal oxides with a similar honeycomb lattice. It is also interesting to note for comparison that due to a large value of $J_2/J_1$=0.15, $Bi_3Mn_4O_{12}(NO_3)$ with a Mn bilayer honeycomb lattice doesn't have an antiferromagnetic long-range order down to 0.4 K [10, 11]. Thus it will be very useful to measure $J_1$ and $J_2$ from inelastic neutron scattering of $Li_2MnO_3$ and compare it with $Bi_3Mn_4O_{12}(NO_3)$. Our latest analysis based on a mean-field model of high-field data [23] suggests that the $J_2/J_1$ ratio is significantly reduced to 0.025 for $Li_2MnO_3$. Thus it is an intriguing question how the small difference in the Mn-Mn distances leads to such a large change in $J_2/J_1$ for the two samples.

Passing comment on the estimation of the Curie-Weiss temperature: One can think of effects of isolated Mn free spins in doped magnetic systems like ours, when estimating the Curie-Weiss temperature as done in $Ba_3NiSb_2O_9$ [24]. However, there are three considerations we made against the idea. First, as we have demonstrated above exchange interaction beyond simple nearest-neighbors are found to be very important for $Li_2MnO_3$. In such a circumstance, it is not trivial to determine the sole contribution to the measured magnetization data due to isolated free Mn spins. Second, there is no low-temperature big increase in the magnetization data as seen in $Ba_3NiSb_2O_9$, which was considered as a sign of such free isolated spin for the latter. Third, that the Curie-Weiss temperature changes its sign for x>0.7 and becomes positive seems to be at variance with the idea of free isolated spins, which otherwise would dominate the magnetic signals for such dilute regime.

Let us now discuss the doping dependence of the transition temperature. Percolation theories for the honeycomb lattice (*d*=2, *z*=3) with only nearest neighbor interaction $J_1$ predict a percolation threshold $p_c$=0.7 of magnetic contents or $x_c$(=1-$p_c$)=0.3 of impurity doping as denoted by hc1 in Fig. 6. This critical threshold ($p_c$) becomes smaller by including farther interactions: $p_c$~0.38 or $x_c$=0.62 for a $J_1$-$J_2$ model (*d*=2, *z*=9) and $p_c$=0.3 or $x_c$=0.7 for a $J_1$-$J_2$-$J_3$ model (*d*=2, *z*=$z_1$+$z_2$+$z_3$=3+6+3=12) denoted by hc1,2,3 in Fig. 6 [12]. The fact that we can trace the antiferromagnetic ordering down to 2 K for x=0.64 indicates that $Li_2(Mn,Ti)O_3$ is more consistent with the $J_1$-$J_2$-$J_3$ model. Here we would like to comment that the previous studies on $Mn_{1-x}Zn_xPS_3$ [13] and $Ba(Ni_{1-x}Mg_x)_2V_2O_8$ [14] with a honeycomb lattice were carried out over a limited x-range up to *x*=0.45 and *x*=0.15, respectively. Given the situations, our studies of $Li_2(Mn,Ti)O_3$ therefore provide a rare opportunity where one can make a direct comparison with the theoretical predictions on the honeycomb lattice over the whole doping range even without changes in the magnetic structure.

In order to make our comparison more relevant, we adopted a mean field result for the nonmagnetic impurity dependence of magnetic transition as below [15]:
$$\frac{T_N(p)}{T_N(1)} = p\text{m}(p)[\text{m}(p) + 1],$$
where *p*(=1-x) is magnetic content, $T_N(p)$ is the transition temperature for a given *p*, and *m(p)* is magnetic moments at zero temperature for a given *p*. Using the experimental data on the ordered moment obtained from the powder neutron diffraction results in Fig. 5, we fitted the data with an empirical formula: $m(p) = m_0(p － p_c)^\alpha$, to obtain that our data can be explained well by the following parameters: $p_c$=0.3, $m_0$ =2.6 $\mu_B$, and $\alpha$=0.28. Two things should be noted. First, we used the data taken at around 7 K for which we have data points for all three samples as shown in Fig. 5. Second, the parameter $\alpha$ is a fitting parameter in our analysis, not to be confused with any critical exponent. Using the formula, we calculated a theoretical curve (solid line in Fig. 6) to find that this theoretical curve describes quite well the reduction of the magnetic transition temperature found in $Li_2$(Mn,Ti)$O_3$. We note that all these observations are made in a system with the same magnetic structure down to the critical concentration.

Finally, we would like to comment that with doping the ratio of $J_2/J_1$ in $Li_2MnO_3$ effectively becomes larger because of the smaller number of nearest neighbors, thus bringing $Li_2$(Mn,Ti)$O_3$ closer to a magnetic instability as seen in $Bi_3Mn_4O_{12}(NO_3)$. This observation then opens an interesting opportunity of tuning the magnetic phase transition near the critical composition by pressure and exploring a possibly quantum critical state for the honeycomb lattice, which is yet an untested ground.

In conclusion, we investigated the physical properties of $Li_2Mn_{1-x}Ti_xO_3$ solid solution system to find that the magnetic structure of $Li_2MnO_3$ remains unchanged upon Ti doping although the antiferromagnetic transition temperature is continuously suppressed with the critical composition of $x_c$=0.7. All our experimental observations are in good agreement with the percolation theories for the *$J_1$-$J_2$-$J_3$* model. This study of the full compositional phase space with a honeycomb lattice offers rare insights into the percolation effects of the honeycomb lattice.

## Acknowledgement

We appreciate K. Balamurugan, Man Duc Le, and Gun Sang Jeon for helpful discussions, and Hongki Min for bringing Ref. 21 to our attention. This work was supported by the Institute for Basic Science (IBS) in Korea.

**Figure Caption**

Figure 1 (Color online) Crystal structure of $Li_2MnO_3$ (C2/m) in (a), (c) and $Li_2TiO_3$ (C2/c) in (b), (d) with projections angle perpendicular (a, b) and parallel (c, d) to the honeycomb plane. $MnO_6$ and $TiO_6$ octahedra share the edges of neighboring octahedra and forms transition metal honeycomb lattices. Honeycomb lattice is stacked along the c-axis after a-axis shifted for $Li_2MnO_3$ and a, b-axis shifted for $Li_2TiO_3$.

Figure 2 (Color online) X-ray diffraction data of $Li_2Mn_{1-x}Ti_xO_3$ refined by using the C2/m (x<0.7) and C2/c (x>0.7) structures, respectively. Superlattice peaks of C2/c clearly visible for x=0.8-1.0 are marked by asterisks in the data for x=0.8. Symbols represent the experimental data and the lines the refinement results with the bottom lines indicating the difference curves and the vertical bars representing the position of nuclear Bragg peaks.

Figure 3 (Color online) (top) Magnetic susceptibility of $Li_2Mn_{1-x}Ti_xO_3$: data for single crystals in the main curve and representative data for powder samples in the inset. (bottom) Doping dependence of Curie-Weiss temperature $\theta_{CW}$ is shown for all the samples: $\theta_{CW}^{Powder}$ for powder samples; $\theta_{CW}^{Single}$ for single crystal samples. The inset shows the plot of the inverse susceptibility for x=0.64 with the symbols (data points) and the line (fitting result).

Figure 4 (Color online) Powder neutron diffraction data are shown of $Li_2Mn_{1-x}Ti_xO_3$ with x=0, 0.25 and 0.5. The data for $Li_2MnO_3$ were reproduced from Ref. 3 for the sake of comparison. For the refinement of the diffraction data, we used the same magnetic structure $\Gamma_{2u}$ with a magnetic propagating vector $k_m$=(0 0 0.5) for all three samples. The vertical bars represent the position of nuclear (upper) and magnetic (lower) Bragg peaks.

Figure 5 (Color online) Temperature dependence of out-of-plane (OP) and in-plane (IP) magnetic components for $Li_2Mn_{1-x}Ti_xO_3$, x=0, 0.25, and 0.5. The lines represent theoretical curve $M(T) = M_0 \left( \dfrac{T_N - T}{T_N} \right)^{\beta}$ using $\beta$=0.28 (see the main text).

Figure 6 (Color online) It shows how the unit cell volume (V) and the antiferromagnetic transition temperature ($T_N$) change with Ti doping with a miscibility gap for x=0.68~0.75. In order to compare the unit cell volume for the same chemical formula, we divided the unit cell volume of the C2/c structure by a factor two in the figure: the dashed line represents the linear Vegard's law. The solid line is our theoretical result using a phenomenological model as discussed in the text with $p_c$ shown for the two $J_1$ (hc1) and $J_1$-$J_2$-$J_3$ (hc1,2,3) models.

Table 1 Summary of refinement results on powder neutron diffraction. Data for $Li_2MnO_3$ were reproduced from Ref. 3 for comparison.

**Figure 1**

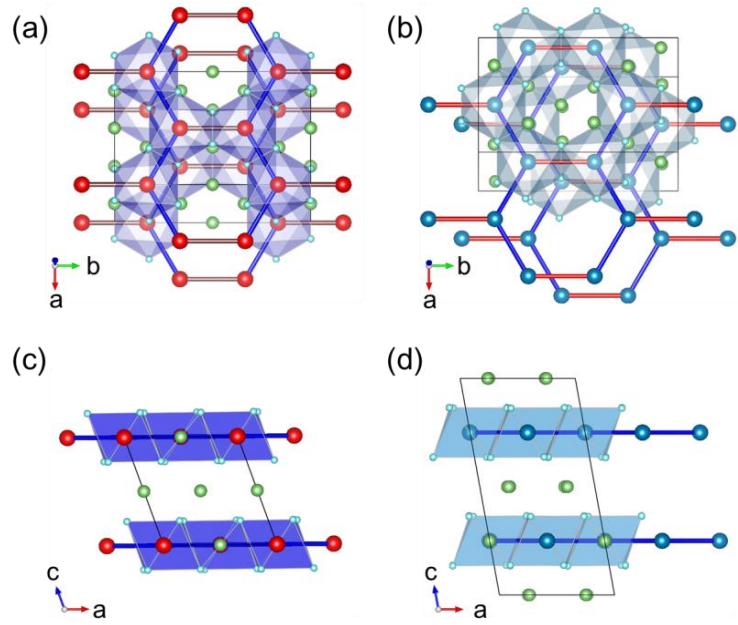

**Figure 2**

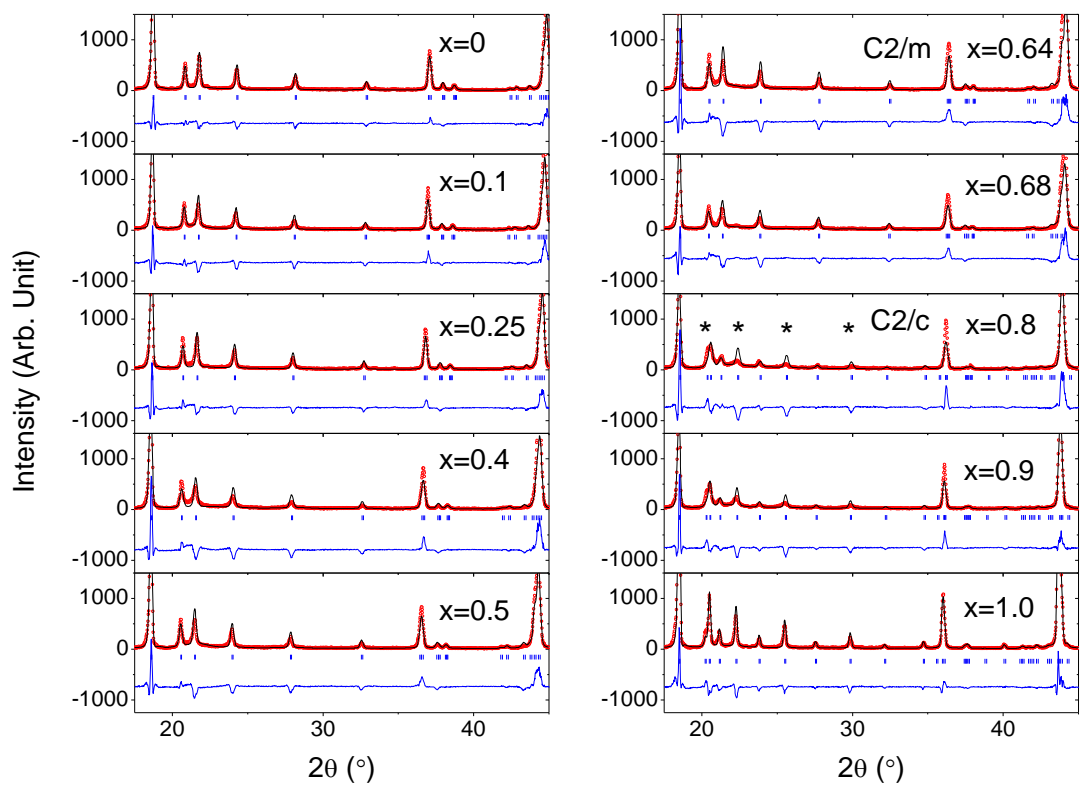

**Figure 3**

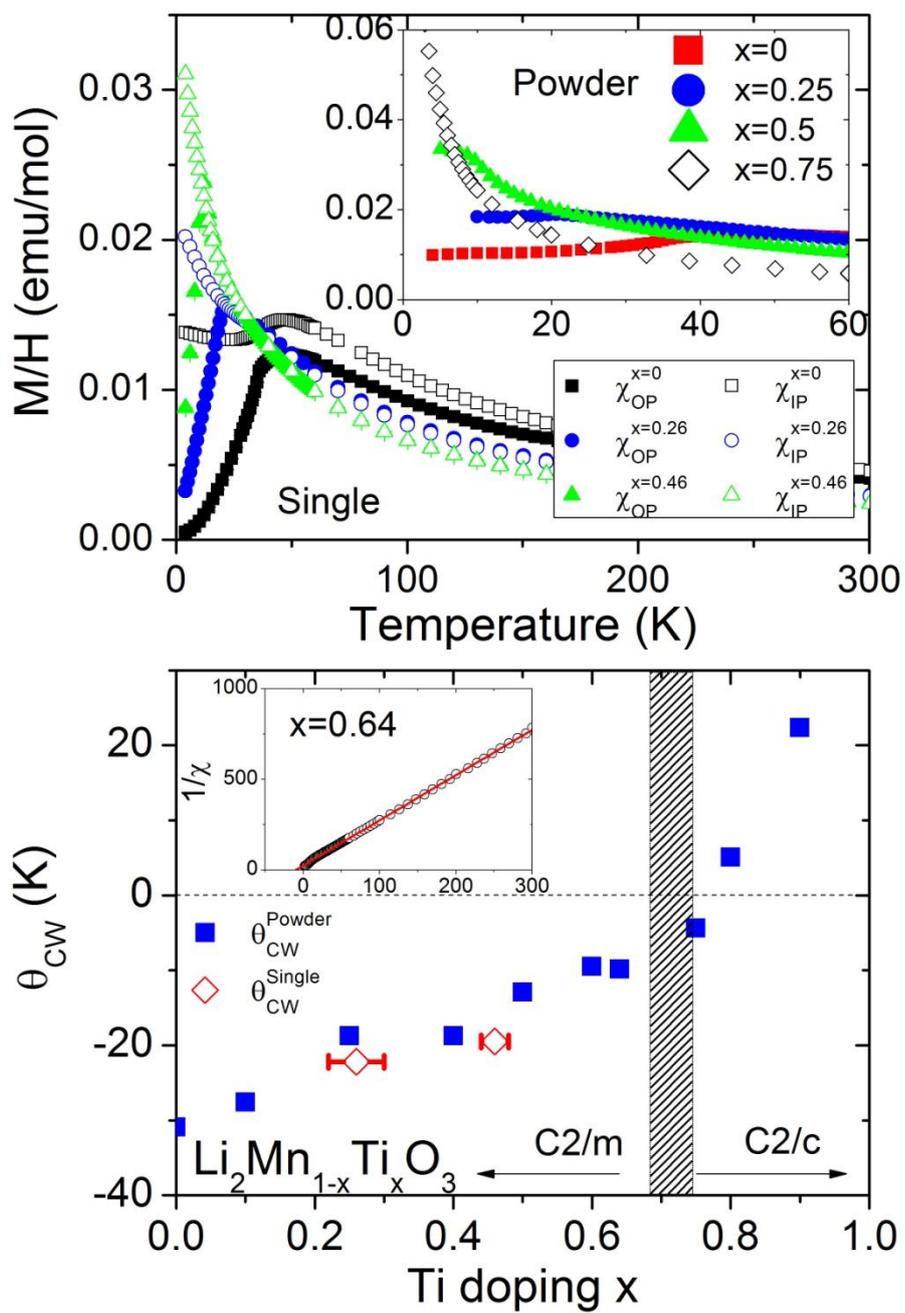

**Figure 4**

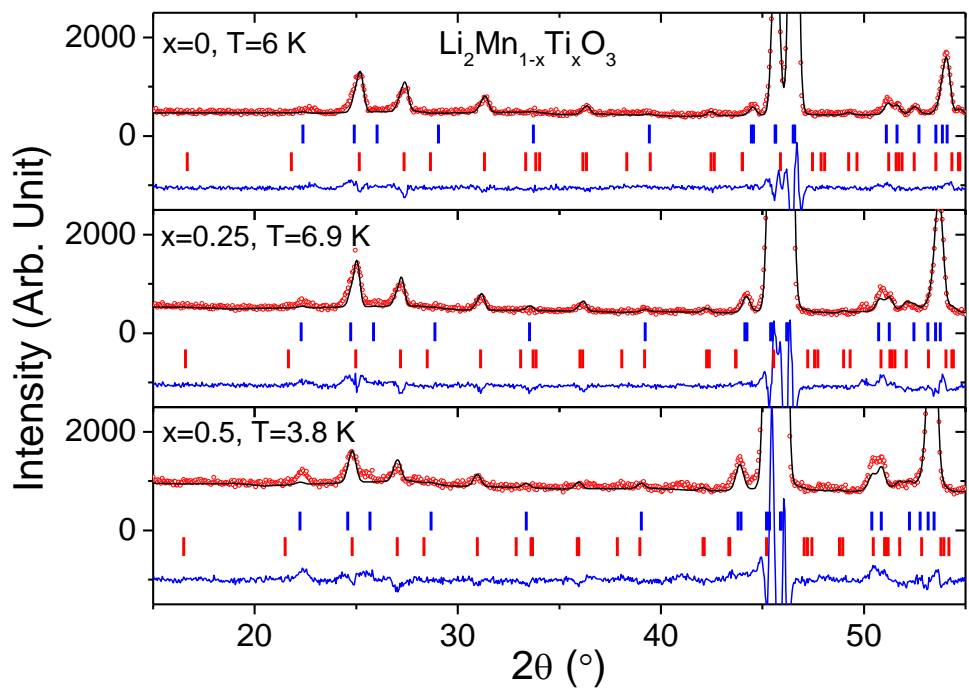

**Figure 5**

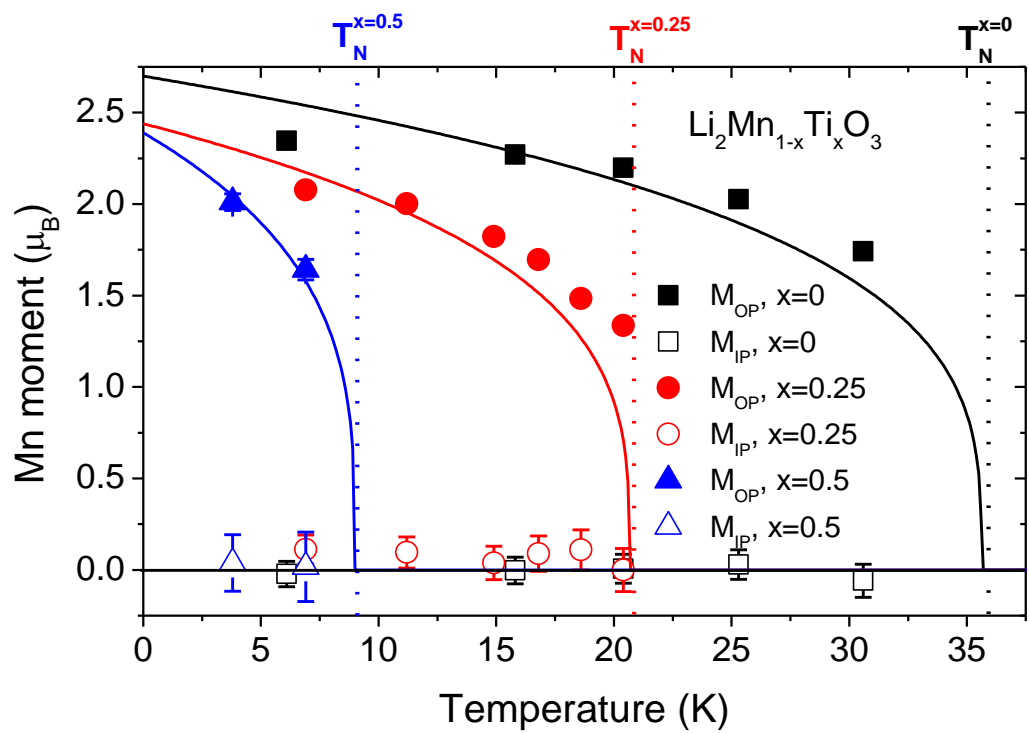

**Figure 6**

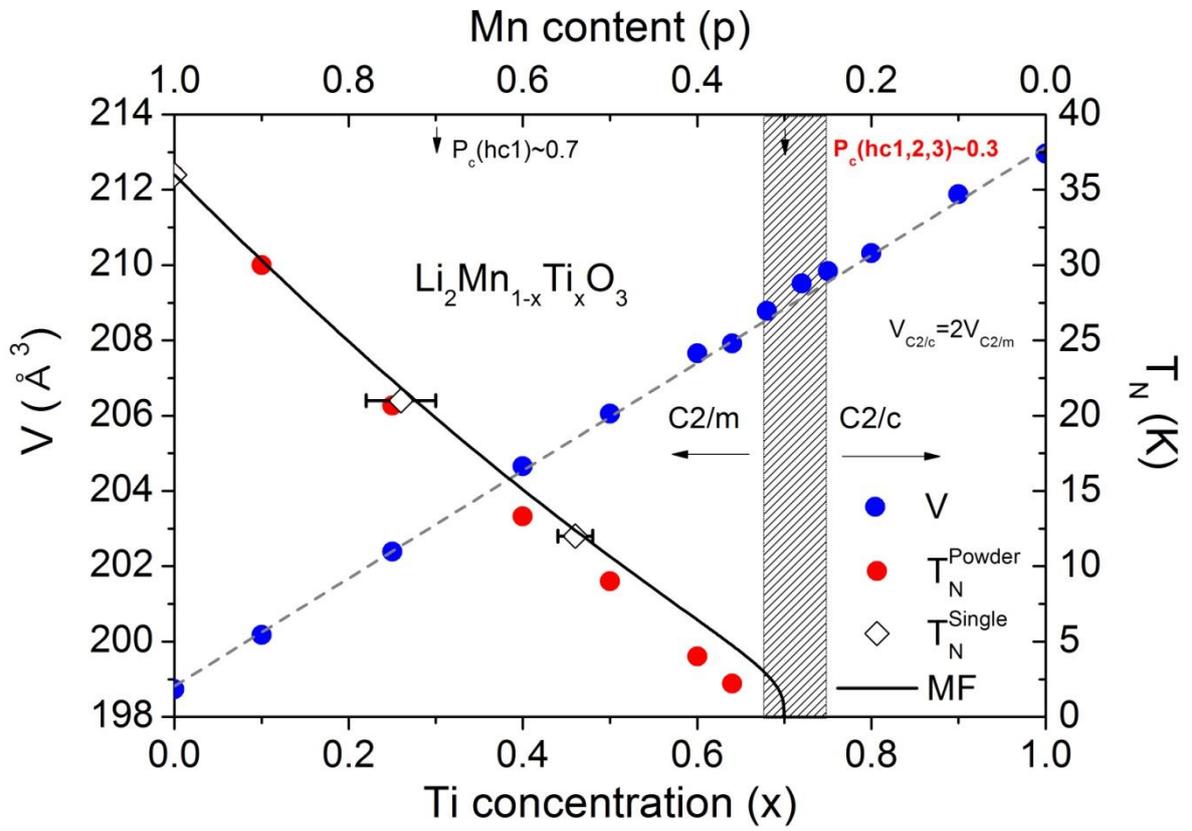

**Table 1** Summary of refinement results on powder neutron diffraction. Data for Li$_2$MnO$_3$ were reproduced from Ref. 3 for comparison.

| | Space group: C2/m (No. 12) $a \neq b \neq c$, $\alpha = \gamma = 90°$, Mn/Ti(4g): (0, y, 0), Li1(2b): (0, 0.5, 0), Li2(2c): (0, 0 0.5), Li3(4h): (0, y, 0.5), O1(4i): (x, 0, z), O2(8i): (x, y, z) | | | | | |
|---|---|---|---|---|---|---|
| | Li$_2$MnO$_3$ | | Li$_2$Mn$_{0.75}$Ti$_{0.25}$O$_3$ | | Li$_2$Mn$_{0.5}$Ti$_{0.5}$O$_3$ | |
| Temperature (K) | 6 K | 60 K | 6.9 K | 35.5 K | 3.8 K | 14.9 K |
| a (Å) | 4.9166(1) | 4.9167(2) | 4.9516(3) | 4.9523(3) | 4.9886(4) | 4.9883(3) |
| b (Å) | 8.5065(2) | 8.5069(2) | 8.5660(5) | 8.5657(5) | 8.6188(6) | 8.6189(5) |
| c (Å) | 5.0117(1) | 5.0099(1) | 5.0309(2) | 5.0309(2) | 5.0495(2) | 5.0496(2) |
| β (°) | 109.376(2) | 109.373(2) | 109.422(3) | 109.426(3) | 109.527(3) | 109.524(3) |
| Volume (Å$^3$) | 197.732(8) | 197.678(9) | 201.248(17) | 201.261(19) | 204.619(22) | 204.617(19) |
| Mn/Ti y | 0.1661(7) | 0.1663(9) | 0.1711(6) | 0.1712(7) | 0.1684(6) | 0.1697(7) |
| Li3 y | 0.6617(20) | 0.6560(20) | 0.6572(13) | 0.6583(14) | 0.6543(14) | 0.6538(14) |
| O1 x | 0.2190(8) | 0.2178(9) | 0.2225(9) | 0.2216(9) | 0.2276(8) | 0.2271(9) |
| O1 z | 0.2260(9) | 0.2253(10) | 0.2221(8) | 0.2217(9) | 0.2287(8) | 0.2286(9) |
| O2 x | 0.2533(5) | 0.2537(6) | 0.2510(6) | 0.2504(6) | 0.2519(5) | 0.2512(6) |
| O2 y | 0.3238(3) | 0.3220(3) | 0.3250(3) | 0.3244(3) | 0.3262(3) | 0.3266(3) |
| O2 z | 0.2231(5) | 0.2237(6) | 0.2286(4) | 0.2292(5) | 0.2268(5) | 0.2272(5) |
| Mn/Ti B$_{iso}$ (Å$^2$) | 0.61(8) | 0.73(9) | 0.24(7) | 0.40(7) | 0.23(6) | 0.03(7) |
| Li B$_{iso}$ (Å$^2$) | 0.93(10) | 0.97(11) | 0.86(9) | 0.45(9) | 1.13(8) | 0.84(9) |
| O B$_{iso}$ (Å$^2$) | 0.60(3) | 0.64(3) | 1.03(3) | 0.85(3) | 1.13(2) | 0.91(3) |
| d$_{TM-TM}^{IP}$ (Å) | 2.826(8) | 2.829(11) | 2.821(3) | 2.820(4) | 2.864(4) | 2.853(4) |
| d$_{TM-TM}^{IP*}$ (Å) | 2.843(4) | 2.841(5) | 2.931(7) | 2.933(8) | 2.903(7) | 2.925(9) |
| M$_x$ (μ$_B$) | 0.80(7) | - | 0.84(8) | - | 0.75(15) | - |
| M$_y$ (μ$_B$) | 0 | - | 0 | - | 0 | - |
| M$_z$ (μ$_B$) | 2.49(2) | - | 2.20(3) | - | 2.13(5) | - |
| M (μ$_B$) | 2.35(2) | - | 2.08(3) | - | 2.01(5) | - |
| θ$_{Spin}$ (°) | 91(2) | - | 87(2) | - | 89(4) | - |
| R$_p$ | 5.26 | 5.28 | 6.85 | 6.95 | 5.73 | 5.89 |
| R$_{wp}$ | 6.80 | 6.84 | 8.97 | 9.32 | 7.56 | 7.73 |
| R$_{exp}$ | 3.86 | 3.99 | 3.66 | 3.65 | 2.72 | 2.70 |
| χ$^2$ | 3.11 | 2.93 | 6.00 | 6.53 | 7.74 | 8.21 |